\documentclass[oneside,12pt]{article}
\usepackage{graphics,color,setspace}
\usepackage{natbib}

\newcommand{\LyA}{{\rm Ly}${\,}\alpha$}
\newcommand{\lya}{\LyA}
\newcommand{\mysection}[1]{\medskip\noindent{\large\bf #1}\par\smallskip}

\topmargin = -\topskip
\advance\topmargin by -\headsep
\begin{document}

\large
\centerline{\bf The Baryon Oscillation Spectroscopic Survey:}
\centerline{\bf Precision Measurements of the Absolute Cosmic Distance Scale}
\normalsize

\bigskip
\centerline
{White Paper for the Astro2010 CFP Science Frontier Panel}

\vspace{1in}

\noindent\underline{Contact Information:}
\vspace{0.1in}

\noindent
David Schlegel,\newline
Lawrence Berkeley National Laboratory,\newline
BOSS Principal Investigator\newline
{\tt djschlegel@lbl.gov}
\vspace{0.1in}

\noindent
Martin White,\newline
U.C. Berkeley,\newline
BOSS Survey Scientist\newline
{\tt mwhite@berkeley.edu}
\vspace{0.1in}

\noindent
Daniel Eisenstein,\newline
University of Arizona,\newline
SDSS-III Director\newline
{\tt deisenstein@as.arizona.edu}

\clearpage
\centerline{\bf ABSTRACT}

BOSS, the Baryon Oscillation Spectroscopic Survey, is a 5-year program to 
measure the absolute cosmic distance scale and expansion rate with 
percent-level precision at redshifts $z<0.7$ and $z \approx 2.5$.  
BOSS uses the ``standard ruler'' provided by baryon acoustic oscillations 
(BAO).  BOSS will achieve a near optimal measurement of the BAO scale at
$z<0.7$, with a redshift survey of 1.5 million luminous galaxies.  It will 
pioneer a new method of BAO measurement at high redshift, using the \lya\ 
forest to 160,000 QSOs in the redshift range $2.1 < z < 3.0$.
The forecast measurement precision for angular diameter distance $d_A$
is 1.0\%, 1.0\%, and 1.5\% at $z=0.35$, 0.6, and 2.5, respectively, and the
forecast precision for the Hubble parameter $H(z)$ is 1.8\%, 1.7\%, and 1.2\%
at the same redshifts.  These measurements will provide powerful constraints
on the nature of dark energy and the curvature of space, complementing the 
constraints obtained from other probes.
BOSS will also provide a superb data set for studying large- and small-scale
clustering, the evolution of massive galaxies and the luminosity function and
clustering of QSOs at $2.3 < z < 6.5$.
BOSS is one of four surveys that comprise SDSS-III (the Sloan Digital Sky
Survey III), a 6-year program that will use highly multiplexed spectrographs
on the 2.5-m Sloan Foundation Telescope to investigate cosmological parameters,
the history and structure of the Milky Way galaxy, and the population of giant
planet systems.

\bigskip

\mysection{1. Introduction}

The acceleration of the expansion of the universe poses the most
profound question in physical science today.  Even the most
prosaic explanations of cosmic acceleration demand the existence
of a pervasive new component of the universe with exotic physical
properties.  More extreme alternatives include extra spatial dimensions
or a breakdown of General Relativity on cosmological scales.
Distinguishing competing hypotheses requires precise measurements
of the cosmic expansion history and the growth of structure.
Among current measurement methods, baryon acoustic oscillations (BAO)
appear to have the lowest level of systematic uncertainty
(Albrecht et al.\ 2006).
Sound waves that propagate in the hot plasma of the early universe
imprint a characteristic scale on the clustering of dark matter,
galaxies, and intergalactic gas.
The length of this ``standard ruler'' can be calculated precisely
using straightforward physics and cosmological parameters that
are well constrained by cosmic microwave background data.
Clustering measurements of the transverse BAO scale yield the
angular diameter distance $d_A$ to the measurement redshift,
while measurements of the line-of-sight BAO scale yield the
Hubble parameter $H(z)$.  The first clear detection of BAO in
the distribution of galaxies was achieved in 2005 by teams
from the Sloan Digital Sky Survey (SDSS; Eisenstein et al.\ 2005)
and the Two Degree Field Galaxy Redshift Survey
(2dFGRS; Cole et al.\ 2005).

The Baryon Oscillation Spectroscopic Survey (BOSS) will
survey the immense volume required to obtain percent-level
measurements of the BAO scale, achieving the precision needed to
provide powerful constraints on the origin of cosmic acceleration.
BOSS will measure redshifts of 1.5 million luminous galaxies and \lya\ 
absorption towards 160,000 high redshift quasars.  
The BOSS galaxy survey will be the definitive low redshift
($z<0.7$) BAO experiment for the foreseeable future because it
covers a large area of sky with spectroscopic redshifts of strongly
clustered tracers, with sampling density sufficient to limit
shot-noise contributions to statistical errors.  Because the
BAO distance scale is anchored in the cosmic microwave background,
the leverage of BAO measurements on dark energy parameters is
(for a given measurement precision) strongest at low redshift.  
BOSS will also lay the groundwork and provide the essential low
redshift comparison point for future baryon oscillation experiments
at higher redshift.  The BOSS \lya\ forest survey will pioneer
a novel method for high-redshift BAO measurements, achieving
valuable constraints on its own and providing a path-finder for
still more powerful surveys that could use this technique
in the future.

BOSS is part of SDSS-III, a six-year program (2008-2014) that will use
existing and new instruments on the 2.5-m Sloan Foundation Telescope to
carry out four spectroscopic surveys on three scientific themes:
dark energy and cosmological parameters;
the structure, dynamics, and chemical evolution of the Milky Way;
and the architecture of planetary systems.\footnote{Because the different
SDSS-III surveys are relevant to different Astro2010 survey panels, we
are providing three separate White Papers, though the general material
on the SDSS is repeated.  A detailed description of SDSS-III is available
at {\tt http://www.sdss3.org/collaboration/description.pdf}.}
All data from SDSS-I (2000-2005) and SDSS-II (2005-2008), fully calibrated
and accessible through efficient data bases, have been made public in a
series of roughly annual data releases, and SDSS-III will continue that
tradition.
SDSS data have supported fundamental work across an extraordinary range of
astronomical disciplines, including the large-scale structure of the universe,
the evolution and clustering of quasars, gravitational lensing, the properties
of galaxies, the members of the Local Group, the structure and stellar
populations of the Milky Way, stellar astrophysics, sub-stellar objects,
and small bodies in the solar system.  A summary of some of the major
scientific contributions of the SDSS to date appears in the Appendix.

By the time of the Astro2010 report, we hope that SDSS-III fund-raising will
be complete.  We are providing SDSS-III White Papers to the Astro2010
committee and panels mainly as information about what we expect to be a major
activity for the first half of the next decade, and about data sets that
will shape the environment for other activities.
We also want to emphasize the value of supporting projects of this scale, 
which may involve public-private partnerships and international collaborations
like the SDSS, and thus the importance of maintaining funds and mechanisms to
support the most meritorious such proposals that may come forward in the next
decade.

\mysection{2. Description of BOSS}

The SDSS telescope still has enormous benefits as a spectroscopic facility
because of its 7 deg${}^2$ field of view.  
BOSS will use upgraded versions of the SDSS
spectrographs with new detectors and gratings to improve red sensitivity and
UV throughput and an increased the number of fibers per field (to 1000 from
640) to increase multiplexing.
These improvements, plus more observation time, allow us to extend SDSS
galaxy redshifts and \lya\ forest measurements to fainter targets.

BOSS will cover $10,000$ deg${}^2$ of high-latitude sky.  Spectroscopic targets
will be selected from SDSS imaging, including 2,000 deg${}^2$ of new imaging
carried out to provide a large contiguous area in the south Galactic cap.
The galaxy sample is pre-selected using color cuts designed to select an
almost stellar mass limited sample of galaxies at $0.4<z<0.7$,
with a subsampling of luminous red galaxies at lower $z$.
The galaxy sample is deeper than the existing SDSS sample
(reaching to $i\simeq 20.0$), probing higher redshift with higher sampling
density and a more representative sample of massive galaxies.
The space density is roughly constant
(at $\bar{n}=3\times 10^{-4}\,h^3\,{\rm Mpc}^{-3})$ to $z\simeq 0.6$
and drops rapidly beyond that, yielding approximately 1.5 million massive
galaxies.
BOSS also targets photometrically selected high-$z$ QSOs near the peak of
the QSO redshift distribution ($2.1< z< 3.0$), which we expect to yield
approximately $20$ QSOs per sq.~deg.\ to $g<22$.

The BOSS galaxy sample will allow a nearly optimal measurement of the acoustic
scale at low $z$.  The larger volume and higher sampling density of BOSS
compared to SDSS-I and II allows better statistical constraints and the
possibility of performing reconstruction (Eisenstein et al.~2007).
By using spectroscopic redshifts BOSS leverages the full power of BAO,
including the ability to measure the expansion rate, and mitigates potential
systematic errors from photometric redshift failures.
By measuring the \lya\ forest in a dense grid of bright QSO spectra, BOSS will
probe the density fluctuations in the IGM at high redshift and measure the
angular diameter distance to, and Hubble parameter at, $z\sim 2.5$.
These observations are important for tightly constraining spatial curvature
and provide key tests of inflation and ``early dark energy'' models.
The dense grid of sightlines in BOSS will enable measurements that are
effectively three-dimensional.  The gain in statistical power is thus much
larger than the `mere' 50-fold increase in QSO numbers would suggest.

Table \ref{tab:fom} presents forecasts of the constraints BOSS would be able
to place on commonly discussed cosmological parameters.
Since the BOSS BAO measurement is independent of the cluster, weak lensing
and supernova methods it allows a cross-check of dark energy conclusions
at a similar signal-to-noise ratio. The combination of two measurements with
similar signal-to-noise ratio more than doubles the FoM because the degeneracy
directions of the constraints are different.

\begin{table}
\begin{center}
\begin{tabular}{lcccccc}
\hline 
Expt. & $h$ & $\Omega_K$ & $w_0$ & $w_p$ & $w_a$ & FoM \\
\hline
BOSS LRG & 0.008 & 0.0028 & 0.089 & 0.032 & 0.366 &  86 \\
{\bf BOSS LRG+QSO} & 0.008 & 0.0019 & 0.076 & 0.029 & 0.279 & {\bf 122} \\
+WL & 0.008 & 0.0017 & 0.068 & 0.026 & 0.227 & 172 \\
+CL & 0.008 & 0.0018 & 0.071 & 0.023 & 0.244 & 177 \\
+SN & 0.006 & 0.0019 & 0.052 & 0.023 & 0.220 & 199 \\
+WL+CL+SN & 0.005 & 0.0016 & 0.046 & 0.018 & 0.164 & 331 \\
\hline
Including Broad-Band Power Information:\\
BOSS LRG+QSO & 0.007 & 0.0015 & 0.065 & 0.016 & 0.240 & 257 \\
+WL+CL+SN & 0.005 & 0.0013 & 0.041 & 0.014 & 0.150 & 479
\end{tabular}
\end{center}
\caption{
Expected 68\% marginalized constraints (absolute, not fractional) from
BOSS on the Hubble parameter ($h$), curvature of space ($\Omega_K$), 
parameterizations
of the equation of state of dark energy ($w_0$, $w_P$, $w_a$),
and the DETF figure
of merit (FoM), which is inversely proportional to the area contained within
the 95\% C.L. of the $w_0-w_a$ plane. All constraints assume the DETF forecasts
for ``Stage II'' experiments, which alone have a FoM of 53, as a prior. 
The first section of the table includes only
the acoustic scale information from the BOSS forecast performance, and shows
the expected improvement from folding in DETF ``Stage III'' constraints from
weak lensing (WL), cluster counts (CL), and supernovae (SN). The second section
combines the forecast performance with BOSS broad-band power
information.
}
\label{tab:fom}
\end{table}

As Table \ref{tab:fom} shows, another remarkable outcome of BOSS will be a
determination of the Hubble constant, $H_0$, with a precision of better
than 1\%, a significant improvement over the HST Key Project determination
{}from a completely different route.
The combination of BAO and SNe will ``invert'' the traditional distance ladder,
tying low redshift tracers to the absolute distance scale calibrated by the
CMB and allowing the absolute luminosities of stellar distance indicators 
to be inferred at the 1\% level from the known value of $H_0$.

Galaxy-galaxy lensing can be used to provide a $S/N\simeq 200$ determination
of the galaxy-mass cross-correlation.  When combined with the observed
amplitude of the galaxy clustering on large scales this provides a percent
level constraint on the amplitude of the mass power spectrum.
On smaller
scales this provides strong constraints on the host dark-matter halos of
massive galaxies, and hence crucial constraints on theories of massive galaxy
formation.
Though it contains similar information, the galaxy-galaxy lensing signal is
complementary to the shear-shear correlation being pursued by other projects
and the availability of spectroscopic redshifts for lenses combined with well
calibrated photometric redshifts for sources allows for some unique advantages
over the deeper shear-shear analyses.
The method can be easily extended once deeper imaging surveys such as
Pan-STARRS or LSST become available over the same area, highlighting the
synergy between these different surveys.

BOSS would be the largest effective volume yet surveyed for large-scale
structure.  Including the effects of shot noise, the galaxy survey will
measure a quarter million Fourier modes at $k<0.2\,h{\rm Mpc}^{-1}$, over
$7\times$ the number in SDSS-II.
The BOSS Ly-$\alpha$ forest spectra would significantly increase the total
line-of-sight distance available for study, providing strong constraints on
small-scale power.
BOSS will thus provide unprecedented constraints on both large- and small-scale
structure, and hence neutrino masses, the running of the spectral index and
warm dark matter.
Further, the study of redshift space distortions will allow us to place percent
level constraints on the growth of structure, which in combination with the
distance measures allows a percent level test of General Relativity on
cosmological scales.

The BOSS data set will be an enormous sample in which to study the evolution
of massive galaxies, traditionally a challenge for theoretical models, which
tend to make these galaxies too massive and too blue.
The combination of SDSS and BOSS data, along with other surveys, will allow
us to trace the process of galaxy formation over half the age of the Universe.
Since activity levels in massive systems are expected to increase at earlier
epochs, such a lever arm in time promises to teach us how the galaxies have
been assembled, and whether the process has resulted in significant energy
transfer to the environment.

Finally, the QSO survey will provide a large and novel sample of less luminous
QSOs at $z=2.5$, near the era of peak QSO activity, which (ironically) is the
most poorly studied redshift range.
This will constrain the faint end of the QSO luminosity function and provide
the best data set for QSO clustering at these redshifts.
The dependence of the clustering on QSO properties, such as luminosity,
provides a novel constraint on the lifetimes and hosts of these rare objects
and can be used to discriminate between the array of currently popular models
for AGN feedback in massive galaxies.
Surveys of high-redshift QSOs have been one of the high impact areas for
SDSS-I.  These objects trace the evolution of early generations of supermassive
black holes, provide tests for QSO formation and AGN evolution, and probe IGM
evolution.
Due to the high impact of the highest redshift quasars and the relatively low
cost of obtaining spectra for these rare objects, BOSS will target higher $z$
quasars as a piggyback program.
By using only a small number of ``extra'' fibers BOSS should approximately
double the number of $z>3.6$ QSOs discovered in SDSS and reach about one
magnitude fainter in the luminosity function.
QSOs at $z>6$ display complete Gunn-Peterson troughs, suggesting that the
reionization phase transition completed just shortly before $z\sim 6$.
BOSS should roughly double the number of rare, $z>6$ QSOs relative
to the SDSS.

The ``guaranteed'' return of BOSS is much tighter constraints on cosmological
parameters, including those describing dark energy, improved constraints from
large- and small-scale structure, new insights into the formation of massive
galaxies, better constraints on QSO demographics near the peak of QSO activity
and a significant increase in the number of known high redshift QSOs.
The ``discovery'' potential is larger still.

\mysection{3. Conclusions}

BOSS will provide a powerful complement to the many proposed large-scale
imaging surveys slated for the next decade
(DES\footnote{\tt www.darkenergysurvey.org};
 Pan-STARRS\footnote{\tt pan-starrs.ifa.hawaii.edu};
 and LSST\footnote{\tt www.lsst.org}) both in terms of data type and scientific
focus.  For dark energy, the imaging surveys largely focus on shear-shear weak
lensing, supernovae and clusters while BOSS focuses on BAO, galaxy-galaxy
lensing and redshift space distortions.  The BOSS spectroscopy will complement
the imaging for a wide range of non-dark-energy science.
BOSS also forms a valuable anchor for future spaced-based spectroscopy
(as for example in one version of JDEM).  

In round numbers, SDSS-III is a \$40 million project, and the
funding is largely in hand thanks to generous support from
the Alfred P. Sloan Foundation, the National Science Foundation,
the Department of Energy, and the Participating Institutions
(including international institutions and participation groups
supported, in some cases, by their own national funding agencies).

SDSS-I and II have 
demonstrated the great value of homogeneous surveys that provide
large, well defined, well calibrated data sets to the astronomical community.
In many cases, such surveys are made possible by novel instrumentation, and
they often require multi-institutional teams to carry them out.
The case for supporting ambitious surveys in the next decade is best made by
considering the contributions of the SDSS to the astronomical breakthroughs of
the {\it current\/} decade, as summarized in the Appendix below.

\clearpage

Funding for SDSS-III has been provided by the Alfred P. Sloan Foundation,
the Participating Institutions, the National Science Foundation, and the
U.S. Department of Energy. The SDSS-III web site is
{\tt http://www.sdss3.org/}.
SDSS-III is managed by the Astrophysical Research Consortium for the
Participating Institutions. The SDSS-III Collaboration is still growing;
at present, the Participating Institutions are the University of Arizona,
the Brazilian Participation Group, University of Cambridge, University of
Florida, the French Participation Group, the German Participation Group,
the Joint Institute for Nuclear Astrophysics, Johns Hopkins University,
Lawrence Berkeley National Laboratory, Max Planck Institute for Astrophysics,
New Mexico State University, New York University, Ohio State University
University of Portsmouth, Princeton University, University of Tokyo,
University of Utah, Vanderbilt University, University of Virginia,
and the University of Washington.

\clearpage
\Large\noindent
{\bf Appendix: The SDSS Legacy}
\normalsize

The SDSS (York et al.\ 2000) is one of the most ambitious and influential
surveys in the history of astronomy.
SDSS-II itself comprised three surveys: the Sloan Legacy Survey completed
the goals of SDSS-I, with imaging of 8,400 square degrees and spectra of
930,000 galaxies and 120,000 quasars; the Sloan Extension for Galactic Understanding and Exploration (SEGUE) obtained 3500 square degrees of additional imaging and spectra of 240,000 stars; and the Sloan Supernova Survey
carried out repeat imaging ($\sim 80$ epochs) of a 300-square degree area,
discovering nearly 500 spectroscopically confirmed Type Ia supernovae for
measuring the cosmic expansion history at redshifts $0.1 < z < 0.4$.
Based on an analysis of highly cited papers, Madrid \& Machetto
(2006, 2009) rated the SDSS as the highest impact astronomical
observatory in 2003, 2004, and 2006 (the latest year analyzed so far).
The final data release from SDSS-II was made public in October, 2008,
so most analyses of the final data sets are yet to come.

The list of extraordinary scientific contributions of the SDSS
includes, in approximately chronological order:
\begin{itemize}
\setlength{\itemsep}{-4pt}
\setlength{\parsep}{0pt}
\item{} {\it The discovery of the most distant quasars,}
tracing the growth of the first supermassive black holes and
probing the epoch of reionization.
\item{} {\it The discovery of large populations of L and T dwarfs,}
providing, together with 2MASS, the main data samples for systematic
study of sub-stellar objects.
\item{} {\it Mapping extended mass distributions around galaxies with weak
gravitational lensing,} demonstrating that dark matter halos extend to several hundred kpc and join smoothly onto the larger scale dark matter distribution.
\item{} {\it Systematic characterization of the galaxy population,}
transforming the study of galaxy properties and the correlations among them
into a precise statistical science, yielding powerful insights
into the physical processes that govern galaxy formation.
\item{} {\it The demonstration of ubiquitous substructure in the outer Milky Way,} present in both kinematics and chemical
compositions, probable signatures of hierarchical buildup of
the stellar halo from smaller components.
\item{} {\it Demonstration of the common origin of dynamical asteroid families,} with distinctive colors indicating similar
composition and space weathering.
\item{} {\it Precision measurement of the luminosity distribution of
quasars,} mapping the rise and fall of quasars and the growth of
the supermassive black holes that power them.
\item{} {\it Precision measurements of large scale galaxy clustering,}
leading to powerful constraints on the matter and energy contents of the Universe and on the nature and origin of the primordial fluctuations
that seeded the growth of cosmic structure.
\item{} {\it Precision measurement of early structure with the Lyman-$\alpha$ forest,} yielding precise constraints on the
clustering of the underlying dark matter distribution
$1.5-3$ Gyr after the big bang.
\item{} {\it Detailed characterization of small and intermediate scale
clustering of galaxies} for classes defined by luminosity, color, and morphology, allowing strong tests of galaxy formation theories
and statistical determination of the relation between galaxies and dark matter halos. \item{} {\it Discovery of many new companions of the Milky Way and Andromeda,}
exceeding the number found in the previous 70 years, and providing
critical new data for understanding galaxy formation in low mass halos.
\item{} {\it Discovery of stars escaping the Galaxy,} ejected by
gravitational interactions with the central black hole, providing
information on the conditions at
the Galactic Center and on the shape, mass, and total extent of
the Galaxy's dark matter halo. \item{} {\it Discovery of acoustic oscillation signatures in the clustering of galaxies,} the first
clear detection of a long-predicted cosmological signal,
opening the door to a new method of cosmological measurement that
is the key to the BOSS survey of SDSS-III.
\item{} {\it Measurements of the clustering of quasars over a wide range
of cosmic time,} providing critical constraints on the dark matter
halos that host active black holes of different luminosities at
different epochs.
\end{itemize}

Half of these achievements were among
the original ``design goals'' of the SDSS, but the other half
were either entirely unanticipated or not
expected to be nearly as exciting or powerful as they turned out to be.
The SDSS and SDSS-II have enabled systematic investigation and
``discovery'' science in nearly equal measure, and we expect that tradition to continue with SDSS-III.


\end{document}